\documentclass[12pt,a4paper]{article}

\usepackage{amssymb,amsmath,amsfonts,eurosym,geometry}
\usepackage{ulem,graphicx,caption,color,setspace}
\usepackage{comment,caption,natbib,pdflscape,subcaption,array,hyperref,amsmath,amsthm,float}
\usepackage{ascmac,bm,tikz,lipsum,sgamevar,natbib,url,enumerate, pgfplots}
\usetikzlibrary{arrows.meta}

\pgfplotsset{compat=1.18}

\usepackage{tcolorbox}
\usepackage{multirow}
\usepackage{lscape}
\usepackage{booktabs}
\usepackage{multirow}
\usepackage{quoting}
\usepackage{colortbl}
\usepackage{subcaption}
\usepackage{titletoc}
\usepgfplotslibrary{groupplots}

\tcbuselibrary{theorems}
\usepackage[T1]{fontenc}
\usepackage{newpxmath}
\usepackage{newpxtext}
\usepackage{hyperref}
\usepackage{diagbox}
\usepackage{threeparttable}

\usetikzlibrary{calc}
\usetikzlibrary{positioning}
\usetikzlibrary{matrix,fit}
\usetikzlibrary{patterns}
\usetikzlibrary{decorations.pathreplacing}

\usepackage[titletoc]{appendix}

\usepackage[immediate]{silence}
\usepackage{sectsty}
\DeactivateWarningFilters[temp]

\allowdisplaybreaks

\theoremstyle{plain}
\newtheorem{theorem}{Theorem}[]

\newtheorem{definition}{Definition}

\newtheorem{proposition}{Proposition}
\newtheorem{lemma}{Lemma}
\newtheorem{remark}{Remark}

\newcommand{\red}[1]{{\color{red} #1}}

\usepackage{natbib}

\makeatletter
\let\@makefntextOring\@makefntext
\def\@makefntext#1{\@makefntextOring{\baselineskip=15pt#1}}
\makeatother

\definecolor{navyblue}{rgb}{0.0, 0.0, 0.5}
\definecolor{green}{rgb}{0.2, 0.5, 0.2}

\hypersetup{
    colorlinks=true,
    citecolor=navyblue,
    linkcolor=navyblue,
    urlcolor=navyblue,
}

\usepackage{etoolbox}
\patchcmd{\thanks}{#1}{\protect\doublespacing}{}{}

\begin{document}

\begin{titlepage}

\title{Inequality-Aware Regulation\thanks{\protect\onehalfspacing I am especially grateful to Matti Liski, Pauli Murto, and Juuso Välimäki for guidance. I am grateful to Piotr Dworczak, Nathan Hancart, Daniel Hauser, Bård Harstad, Johannes Hörner, Andreas Kleiner, Eero Mäenpää, Tuomas Markkula, Léopold Monjoie, Juan-Pablo Montero, Jukka Pirttilä, Markus Reisinger, Julia Salmi, François Salanié, Johannes Schneider, Ryo Shirakawa, Peter Norman Sørensen, Egor Starkov, and audiences at the University of Bonn, at the University of Copenhagen, at Copenhagen Business School, at Helsinki GSE, at Inequality-aware Market Design Days (Warsaw) and IMD Online Seminar, at JKU Linz, at the Nordic Economic Theory workshop 2025 (Copenhagen) and NET Online Seminar, at the University of Oslo, and at the Toulouse School of Economics for their insightful comments. The project has received funding from the European Research Council (ERC) under the European Union’s Horizon 2020 research and innovation programme (101040122---IMD and 101055353---EXPLOIT). }} \author{ Mikael Mäkimattila\thanks{University of Oslo and GRAPE. Email: m.p.e.makimattila@econ.uio.no.} }

\date{ \today\\[.5em] \small }

\maketitle

\enlargethispage{1.0\baselineskip}

\begin{abstract}
I study regulation of a screening monopolist when the regulator cares about the distribution of surplus across heterogeneous consumers and the firm.  Consumers privately know their willingness to pay for quality, and the firm privately knows its production costs. Regulatory contracts specify quality--price options that the firm must offer. I show that when the regulator gives greater weight to consumers with lower willingness to pay, \emph{baseline regulation} is optimal: the regulator requires affordable basic options but leaves the firm free to design and price premium options. Under strong consumer-side redistributive concerns, a single required basic option suffices. The regulator uses the firm's incentive to extract surplus from premium buyers to support low regulated prices for basic options.
\vspace{0in}
\end{abstract}

\setcounter{page}{0}
\thispagestyle{empty}
\end{titlepage}

\newcolumntype{C}[1]{>{\centering\arraybackslash}p{#1}}

\onehalfspacing

\setlength{\abovedisplayskip}{5pt}
\setlength{\belowdisplayskip}{5pt}

\section{Introduction} \label{sec: intro}

Market power and distributional concerns are central motivations for regulatory intervention in private markets. These two rationales for intervention often interact: a firm exercising market power designs its offering strategically, affecting the distribution of surplus both among consumers and across market sides. For example, concentration in the private health care industry---a major policy concern in many countries---has implications for the quality of care available to the poor and the mark-ups paid by the rich.

The theory of regulation has developed a rich understanding of how to control market power when firms know their operating environment better than the regulator. However, the standard approach evaluates consumer welfare through aggregate surplus, largely abstracting from how that surplus is distributed among consumers. I develop a theory of optimal regulation that places consumer heterogeneity and redistributive concerns at the center of the analysis. This brings together two interacting design problems: the regulator screens a privately informed firm with a regulatory mechanism that  shapes how the firm screens consumers and distributes surplus among them.

I consider the following model. As in a standard model of second-degree price discrimination, a profit-maximizing firm offers a menu of qualities and prices, and consumers choose from the menu according to their privately known willingness to pay (WTP) for quality. A regulatory contract specifies which quality–price options the firm must make available to consumers; the firm may also offer additional options not required by regulation. As in a standard model of regulation, the firm's production costs are privately known, and the regulator may offer a choice of contracts to screen the firm. The regulator maximizes a weighted utilitarian objective, placing greater weight on consumers on average than on the firm and weakly higher weights on consumers with lower WTP. This captures applications in which low WTP tends to reflect limited economic resources.

My main result shows that optimal regulatory contracts take the form of baseline regulation. The regulator requires the firm to offer affordable basic-quality options consumed by low-WTP consumers, while leaving the firm free to design and price its premium options. When consumer-side redistributive concerns are strong, it suffices to require a single basic option, chosen by a range of low-WTP consumers. The regulator may screen the firm through the choice of such contracts.

Baseline regulation uses the firm's pricing incentives to support redistribution. The surplus the firm extracts from premium buyers enables the regulator to require affordable basic options for consumers whose utility  it values more. These options provide utility guarantees that also shape the firm's additional offers. Even consumers choosing premium options benefit relative to laissez-faire: their ability to choose a basic option limits what the firm can charge them.

Basic qualities lie between the unregulated monopoly and surplus-maximizing levels. Two redistributive considerations favor limiting basic quality. First, lower basic quality makes these options less attractive to high-WTP consumers, who value quality more. This permits higher premium prices that support lower regulated prices for basic options. Second, since production costs are privately known, requiring higher basic quality entails either leaving more information rents to low-cost firms---shifting surplus from consumers to the firm---or accepting a greater probability of shutdown.

Strong consumer-side redistributive concerns can make optimal regulation especially simple: a single basic option suffices. Such concerns make it optimal to pool a range of low-WTP consumers on a common basic option, as separating them would create rents that the regulator values relatively little. For higher-WTP consumers, strong redistributive concern makes the regulator focus on the revenue they generate to support this basic option, so that it can rely on the firm's pricing incentives to supply the premium menu.

Policies resembling baseline regulation are used in a range of markets.  In broadband, for example, New York's Affordable Broadband Act requires providers to offer an affordable basic-speed product  but allow pricing flexibility for higher-speed services. Providers can choose between a lower-speed basic service under a lower price cap and a faster basic service under a higher cap. This paper shows how such a choice among basic quality--price requirements can optimally screen firms with private information.

Similar policies exist in health and social care. German regulation requires private health insurers to offer standardized basic coverage at a capped premium, but they retain flexibility in pricing more comprehensive policies (e.g.,  extensive dental coverage).\footnote{ The \emph{Basistarif} caps the premium on standard coverage. See the German Federal Ministry of Health, \emph{Basistarif in der privaten Krankenversicherung}. } In residential aged care in countries such as New Zealand, providers must supply prescribed standard services at regulated fees, but they retain greater flexibility to design and price optional services such as larger rooms or rooms with private bathrooms.\footnote{ See the New Zealand Government, \emph{Paying for Residential Care}.}

When quality is interpreted as quantity, corresponding policies make low levels of consumption available at an affordable price. Increasing-block electricity and water tariffs, for example, charge lower marginal prices for low levels of consumption than for additional use.\footnote{See \citet{borenstein2012redistributional} on nonlinear electricity pricing in California and \citet{olmstead2007water} on water pricing in North America.}

This paper provides a distributional rationale for such baseline regulation. It also highlights the trade-offs in designing the basic options in applications. Raising basic quality constrains premium markups and therefore the scope to require a low basic price. The firm’s private information about costs introduces a further tension: requiring higher basic quality may entail leaving greater rents to the firm to maintain provision, or accepting shutdown risk.

In Section~\ref{sec: discussion}, I study two variations of the model. In the first variation, I allow the regulator to fully control the terms of trade, including restricting side contracting in which the firm offers consumers more attractive options. Such restrictions can help the regulator limit the firm's information rents: setting a tight cap on premium quality under a high-cost report makes that contract less attractive to a low-cost firm. In the second variation, I consider welfare weights that increase with WTP, which may be natural, for example, if WTP reflects needs rather than ability to pay. Optimal regulation can then take a different form: for example, it can become optimal to require the firm to offer a single subsidized premium option chosen by high-WTP consumers, while leaving the firm free to design and price lower-quality options.

\paragraph{Related literature.}
Theory of regulation studies how to control market power when firms are better informed than the regulator about their operating environment \citep[see][for a survey]{armstrong2007recent}. In the canonical  model of \citet{baron1982regulating}, the regulator balances efficiency against the rents needed to elicit the firm's privately known costs. \citet{amador2022regulating} analyze optimal regulation without transfers. This literature typically evaluates consumer welfare through aggregate consumer surplus. I model the consumer side more richly: consumers select from a menu according to their heterogeneous preferences, and the regulator assigns welfare weights that vary across consumers.

A related strand studies regulation of a screening monopolist with limited regulatory instruments, mostly abstracting from the firm's private information and redistributive preferences among consumers.  \citet{besanko1987monopoly} analyze price caps and quality floors, \citet{krishna1990protection} studies specific and ad valorem taxes and subsidies, \citet{corts1995regulation} examines how a cap on basic-service prices affects pricing and bundling, \citet{schlom2024essays} studies price distribution regulation, and \citet{xia2025regulating} studies optimal regulation of quality and coverage.

The paper contributes to recent work on inequality-aware market design, which characterizes the optimal allocation of goods under redistributive objectives, allowing for a broad range of instruments, including rationing \citep{dworczak2021redistribution,akbarpour2024redistributive,monjoie2026redistributive,kang2026topping}. This literature typically abstracts from suppliers' market power, whereas here redistribution operates through a privately informed firm that designs the menu for consumers. Relatedly, there is work that examines how firms' contracting decisions shape optimal income taxation: \citet{da2019optimal} study noncompetitive labor markets, while \citet{stantcheva2014optimal} studies firms that competitively screen workers through nonlinear compensation contracts.

The connection between consumer-side inequality and market power has received surprisingly limited attention. \citet{leisten2024economic} studies how wealth inequality affects pricing incentives and develops an inequality-adjusted measure of consumer surplus. \citet{augias2026redistribution} characterize optimal redistributive segmentation of a monopoly market. There is also empirical work that examines distributional implications of market power \citep{herkenhoff2025who,dopper2025rising}.

The paper also relates to work in which a designer shapes subsequent contracting by another principal. \citet{dworczak2026mechanism} study the design of property rights, which determine an agent's outside options in a subsequent screening mechanism. \citet{bhaskar2023regulation} study regulation of insurance menus when insurers possess information about risks that is unavailable to the regulator. The different applications motivate different assumptions about the parties' objectives and the designer's control over downstream contracting.

\section{Set-up}\label{sec: model}

\paragraph{Market environment.} As in a standard model of second-degree price discrimination \citep{mussa1978monopoly,maskin1984monopoly}, a profit-maximizing firm offers a menu of quality--price options $(q,p)\in\mathbb R_+\times\mathbb R$ to a unit mass of consumers. Quality may also be interpreted as quantity. A consumer with willingness-to-pay (WTP) type $\theta$ obtains utility $\theta v(q)-p$. She chooses an option from the firm's menu or declines to purchase, obtaining zero.

Consumer types $\theta$ are privately observed and distributed according to $F$ on $\Theta=[\underline\theta,\overline\theta]$, where $0<\underline\theta<\overline\theta<\infty$. The density $f$ is strictly positive, continuously differentiable, and log-concave. The function $v:\mathbb R_+\to\mathbb R_+$ is continuous and unbounded, with $v(0)=0$ and $v'(q)>0$, $v''(q)<0$ on $(0,\infty)$, and the Inada conditions $ \lim_{q\downarrow0}v'(q)=\infty$ and $ \lim_{q\to\infty}v'(q)=0$.

Serving a consumer with quality $q$ costs $\beta q$.\footnote{Non-linear production costs can also be accommodated by a change of units. If producing ``physical quality'' $\tilde q$ costs $\beta c(\tilde q)$, where $c$ is continuous, strictly increasing, and unbounded with $c(0)=0$, set $q=c(\tilde q)$ and express consumer benefits in these units. My analysis then applies (provided that the transformed benefit function satisfies the assumptions on $v$).} As in a standard model of regulation \citep{baron1982regulating}, the firm privately observes its marginal cost type $\beta\in(\underline\beta,\overline\beta)$, drawn independently of consumer types from $G$, where $0<\underline\beta<\overline\beta<\infty$. The density $g$ is strictly positive, continuously differentiable, and log-concave. The surplus-maximizing quality is therefore
\[
    q^e(\theta,\beta)
    =(v')^{-1}\!\left(\frac{\beta}{\theta}\right).
\]

\paragraph{Redistributive preferences.}
The regulator attaches welfare weight $\omega(\theta)$ to the utility of consumer type $\theta$ and weight $\omega_F$ to firm profit. These weights measure the social value of an additional unit of money received by each participant.

Unlike standard models of regulation, I allow the consumer weight $\omega$ to depend on WTP. The weight $\omega(\theta)$ can be interpreted as the expected social welfare weight conditional on WTP. I assume that $\omega$ is nonnegative and weakly decreasing. Decreasing weights are natural in applications where lower WTP largely reflects limited economic resources rather than tastes or needs alone.\footnote{ Part of the literature on mechanism design and redistribution explicitly models heterogeneous marginal utilities of money and assumes the designer maximizes the sum of consumers' utilities. When utilities are expressed in monetary units, each consumer's marginal utility of money becomes her effective welfare weight. See, e.g., \cite{dworczak2021redistribution}; the approach dates back at least to \cite{weitzman1977price}.  A similar approach could be used here as well. Another approach, common in the public finance literature, applies a concave transformation to individual utilities. The weighted objective can be interpreted as a local approximation to such a social welfare function. }

I also assume that the regulator places greater weight on consumers on average than on the firm:
\[
    0\leq\omega_F<\overline\omega
       :=\int_\Theta\omega(\theta)\,dF(\theta).
\]

Writing $U(\theta,\beta)$ for equilibrium consumer utility and $\Pi(\beta)$ for firm profit, the regulator maximizes
\begin{equation}\label{eq: setup welfare}
    \int_{\underline\beta}^{\overline\beta}
       \left[
           \int_\Theta\omega(\theta)U(\theta,\beta)\,dF(\theta)
           +\omega_F\Pi(\beta)
       \right]dG(\beta).
\end{equation}

\paragraph{Regulation.}
A regulatory contract specifies a set of quality--price options that the firm must offer to consumers if it operates; the firm may also offer additional options. This reflects the idea that verifying the availability of required options may be easier than preventing the firm and consumers from agreeing on terms they both prefer. The regulator may offer a menu of contracts to screen the firm. In Section~\ref{sec: flexible}, I consider fully flexible regulation, under which the regulator can also prohibit side contracting.

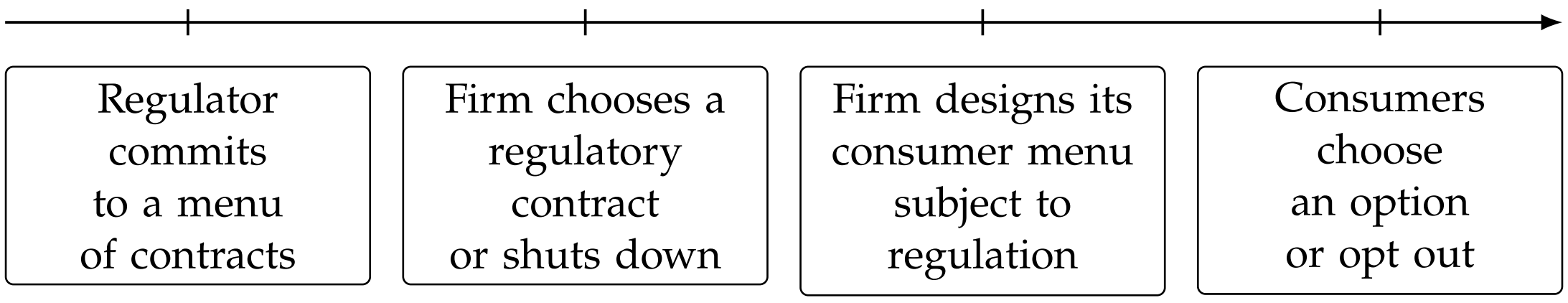
\begin{figure}[t]
\centering
    \begin{tikzpicture}[
        x=\linewidth,
        y=1cm,
        stage/.style={
            draw,
            rounded corners=2pt,
            line width=0.5pt,
            fill=white,
            align=center,
            font=\small,
            text width=0.21\linewidth,
            minimum height=2cm,
            inner sep=4pt,
            anchor=north
        }
    ]
        \draw[-{Latex[length=2mm]}, line width=0.7pt]
            (0.01,0) -- (0.99,0);
        \foreach \x in {0.125,0.375,0.625,0.875}
            \draw[line width=0.7pt] (\x,-0.12) -- (\x,0.12);

        \node[stage] at (0.125,-0.4)
            {Regulator commits\\to a menu\\of contracts};
        \node[stage] at (0.375,-0.4)
            {Firm chooses a\\regulatory contract\\or shuts down};
        \node[stage] at (0.625,-0.4)
            {Firm designs its\\consumer menu\\subject to regulation};
        \node[stage] at (0.875,-0.4)
            {Consumers choose\\an option\\or opt out};
    \end{tikzpicture}
\caption{Timeline of the regulatory game.}
\label{fig:timeline}
\end{figure}

The timing is illustrated in Figure~\ref{fig:timeline}. The regulator first designs a menu of regulatory contracts. Knowing its cost type $\beta$, the firm chooses a contract or shuts down, in which case all parties receive zero payoffs. An operating firm then designs its consumer menu, which must include the quality--price options required by regulation but may also include additional options not required by regulation. Finally, consumers choose an option from the firm's menu or opt out, given their privately known WTP types $\theta$. I study perfect Bayesian equilibria of the game.

\paragraph{Scope of regulation.}
The paper takes a partial-equilibrium perspective: a single market is modeled, and the distribution of WTP and welfare weights are taken as given. The model therefore abstracts from other redistributive instruments such as income taxation. One justification is political feasibility: policymakers regulating a specific market may lack authority or political capital in other policy domains. Another is that governments may have market-specific redistributive objectives \citep[][calls this “specific egalitarianism”]{tobin1970limiting}. Moreover, recent work shows that redistributive interventions in goods markets remain optimal even when income taxation is optimally designed \citep[e.g.,][]{doligalski2025optimal, ahlvik2024pigouvian}.

Within the market, both market power and redistribution are addressed by regulating the firm's menu. Another instrument for consumer-side redistribution might be to subsidize consumers' purchases. A lower regulated price may, however, reproduce the same outcome as a purchase subsidy financed by a tax on the firm. Even if consumer subsidies were used, they should be designed jointly with firm-side regulation, allowing subsidies to depend on the menu the firm offers or the regulatory contract it selects. It is therefore not generally optimal to address inequality and market power independently.

\subsection{Preliminary analysis}\label{sec:preliminaries}

\paragraph{Consumers' problem.}
Write $q(\theta,\beta)$ and $p(\theta,\beta)$ for the equilibrium quality and payment of consumer type $\theta$ when the firm's cost type is $\beta$. A consumer who does not purchase has $q=p=0$.

By a Myersonian argument, consumer incentive compatibility is equivalent to quality being nondecreasing in $\theta$ and utility satisfying the envelope formula
\begin{equation}\label{eq: setup consumer envelope}
    U(\theta,\beta)
       =U_0(\beta)
          +\int_{\underline\theta}^{\theta}
               v(q(t,\beta))\,dt,
    \qquad
    U_0(\beta):=U(\underline\theta,\beta).
\end{equation}
Consumer participation is equivalent to $U_0(\beta)\geq0$. Alternatively, consumer allocation is pinned down by  \(U(\cdot,\beta)\): incentive compatibility is equivalent to  \(U(\cdot,\beta)\) being nondecreasing and convex, and the slope determines quality through \(U_\theta=v(q)\) almost everywhere.

\paragraph{Firm's problem.}

Let $\mathcal P(\beta)$ denote the set of required options under the equilibrium contract of type $\beta$. Conditional on the firm operating, this contract guarantees consumers the utilities
\begin{equation}\label{eq: required utility}
    \underline U(\theta,\beta)
       :=\max\left\{0,\,
           \sup_{(q,p)\in\mathcal P(\beta)}
               [\theta v(q)-p]\right\}.
\end{equation}
Additional options cannot reduce any consumer's utility below this guarantee. Conversely, any incentive-compatible utility function above the guarantee can be implemented while retaining the required options: adding them to an implementing menu leaves consumers' maximal utilities unchanged. Two required menus that generate the same guarantees are therefore equivalent for the firm's incentives.

After selecting contract $\hat\beta$, a firm with cost $\beta$ chooses the most profitable nondecreasing convex utility function above this guarantee. Using the consumer envelope and integrating by parts gives its maximal profit:
\begin{equation}\label{eq: firm continuation value}
    \pi(\hat\beta;\beta)
       =\sup_{\substack{
              u\ \text{nondecreasing and convex}\\
              u\geq\underline U(\cdot,\hat\beta)}}
          \left\{
             \int_\Theta
                [\phi(\theta)u'(\theta)
                   -\beta v^{-1}(u'(\theta))]\,dF(\theta)
             -u(\underline\theta)
          \right\},
\end{equation}
where $\phi(\theta):=\theta-[1-F(\theta)]/f(\theta)$ is the usual virtual value.

Without regulation, the utility guarantee is zero. The firm leaves the lowest consumer zero utility and chooses qualities to maximize virtual surplus. Log-concavity of $f$ makes the virtual value $\phi$ strictly increasing, so pointwise maximization gives a nondecreasing monopoly allocation:
\begin{equation}\label{eq: monopoly quality}
    q^m(\theta,\beta)=
    \begin{cases}
       0, & \phi(\theta)\leq0,\\[3pt]
       (v')^{-1}\!\left(\dfrac{\beta}{\phi(\theta)}\right),
          & \phi(\theta)>0.
    \end{cases}
\end{equation}

The firm's freedom to add options makes this benchmark a lower bound for quality under regulation: $q(\theta,\beta)\geq q^m(\theta,\beta)$ whenever the firm operates. Otherwise, the firm could increase profit by replacing $q$ by $\max\{q,q^m\}$ while keeping $U_0$ fixed. This modification would preserve consumer incentive compatibility and weakly raise every consumer's utility, so it would respect the regulatory guarantee.

\begin{figure}[t]
\centering
   \begin{tikzpicture}
        \pgfmathdeclarefunction{firmHighPrimitive}{1}{\pgfmathparse{4*(16/3*(#1-1)^3+8*(#1-1)^2
                -12*(#1-1)-8*ln(#1-1)-4/(#1-1))/81}}
        \pgfmathsetmacro{\thetaL}{(18.95-sqrt(57.5025))/8}
        \pgfmathsetmacro{\thetaH}{(10.95+sqrt(40.7025))/8}
        \pgfmathsetmacro{\vB}{1.21}
        \pgfmathsetmacro{\pB}{1.35}
        \begin{axis}[
            width=0.80\textwidth, height=7cm,
            axis lines=left,
            xlabel={Consumer type $\theta$}, ylabel={Utility},
            xmin=0.96, xmax=3.05, ymin=0, ymax=3.45,
            xtick={1,\thetaL,\thetaH,3},
            xticklabels={$\underline\theta$,$\theta_L$,
                         $\theta_H$,$\overline\theta$},
            ytick=\empty,
            grid=none, tick align=outside,
            legend columns=1,
            legend style={at={(0.03,0.97)}, anchor=north west,
                draw=none, fill=none, font=\small,
                row sep=2pt},
            legend cell align=left,
            samples=201
        ]
            \addplot[black!55, ultra thick, dashed, domain=1:3]
                {firmHighPrimitive(max(x,1.5))-firmHighPrimitive(1.5)};
            \addlegendentry{Unregulated monopoly}

            \addplot[blue!75!black, ultra thick, domain=1:\thetaL]
                {(\vB*\thetaL-\pB)
                    -\vB*(\thetaL-1)/3
                       *(1-((x-1)/(\thetaL-1))^3)};
            \addlegendentry{Firm's response to regulation}
            \addplot[blue!75!black, ultra thick, forget plot,
                domain=\thetaL:\thetaH] {\vB*x-\pB};
            \addplot[blue!75!black, ultra thick, forget plot,
                domain=\thetaH:3]
                {(\vB*\thetaH-\pB)
                    +firmHighPrimitive(x)-firmHighPrimitive(\thetaH)};

            \addplot[black, very thick, densely dotted, domain=1:3]
                {max(0,\vB*x-\pB)};
            \addlegendentry{Utility guarantee}

            \draw[black!35, thin, densely dashed]
                (axis cs:\thetaL,0) -- (axis cs:\thetaL,{(\vB*\thetaL-\pB)});
            \draw[black!35, thin, densely dashed]
                (axis cs:\thetaH,0) -- (axis cs:\thetaH,{(\vB*\thetaH-\pB)});
            \addplot[blue!75!black, only marks, mark=*, mark size=2pt,
                forget plot] coordinates
                {(\thetaL,{(\vB*\thetaL-\pB)})
                 (\thetaH,{(\vB*\thetaH-\pB)})};
        \end{axis}
    \end{tikzpicture}
\caption{\small  A required option and the firm's response. The dashed curve shows consumer utility under unregulated monopoly, and the dotted curve shows the utility guarantee provided by a single option that regulation requires. The solid curve shows utility under the firm's optimal response. Consumers with $\theta\in[\theta_L,\theta_H]$ choose the required option, while lower and higher types choose additional options offered by the firm.    }
\label{fig: firm response}
\end{figure}
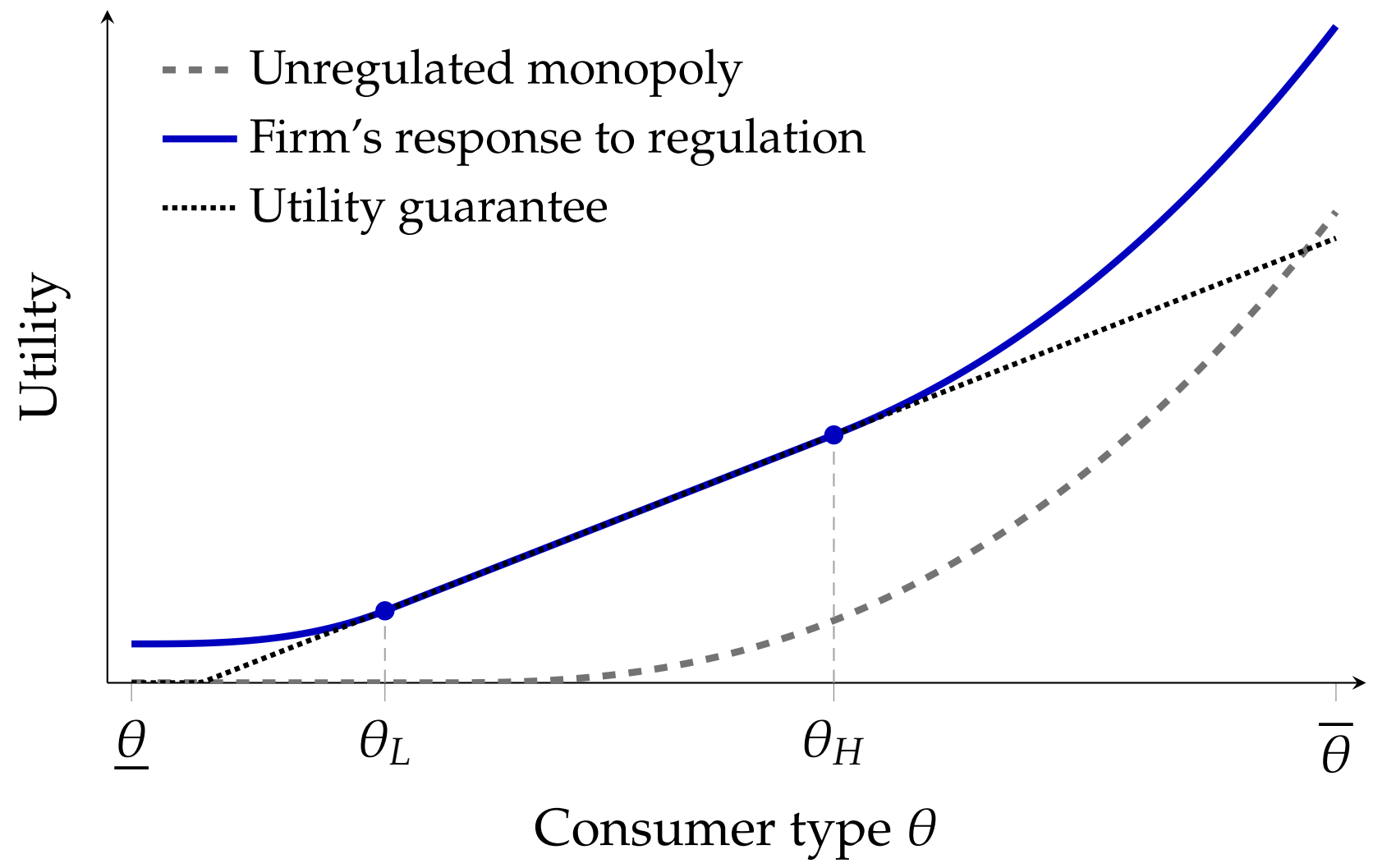

Figure~\ref{fig: firm response} illustrates the firm's response to a single required option. In this example, only consumers with intermediate WTP choose this option: the firm supplements it with cheaper, lower-quality options for low types and premium options for high types. High-WTP consumers receive monopoly qualities, so their utility curve has the same slope as under unregulated monopoly. They nevertheless pay lower prices because the possibility of choosing the required option constrains the pricing of premium options.

\paragraph{Regulator's problem.}
We may focus on equilibria in which firms operate below a cost cutoff $b$. The firm's rents have a redistributive cost because $\overline\omega>\omega_F$. An optimal mechanism therefore leaves zero profit at the operating cut-off $b$. The firm's envelope condition then gives
\begin{equation}\label{eq: minimum firm rent}
    \Pi(\beta)
       =\int_\beta^{\overline\beta}\int_\Theta
           q(\theta,s)\,dF(\theta)\,ds,
\end{equation}
where $q(\theta,s)=0$ whenever type $s$ shuts down. Lower-cost firms retain information rents determined by the provision of higher-cost firms.

The envelope conditions allow the regulator's payoff to be expressed in terms of qualities. Define the \emph{redistributive virtual value} and the \emph{virtual cost} by
\begin{equation}\label{eq: virtual primitives}
\begin{aligned}
    \psi(\theta)
       &:=\phi(\theta)+
          \frac{\displaystyle
              \int_\theta^{\overline\theta}
                   \omega(t)\,dF(t)}
               {\overline\omega f(\theta)},\\
    \widetilde\beta(\beta)
       &:=\beta+
          \left(1-\frac{\omega_F}{\overline\omega}\right)
          \frac{G(\beta)}{g(\beta)}.
\end{aligned}
\end{equation}
The usual virtual value $\phi$ accounts for the revenue lost through consumer information rents. The regulator also values these rents, so $\psi$ adds their contribution to welfare. Decreasing consumer weights imply $\phi\leq\psi\leq\theta$; with equal consumer weights, $\psi(\theta)=\theta$.

The virtual cost $\widetilde\beta$ accounts for information rents on the firm side. Greater quality provision at cost $\beta$ raises the rents that must be left to all lower-cost firms. The adjustment from $\beta$ to $\widetilde\beta$ values these rents at the redistributive loss from shifting surplus from consumers to the firm.

We obtain the following relaxed allocation problem that sets aside consumer participation and, partly, the firm's incentive compatibility:
\begin{equation}\tag{R}\label{eq: relaxed quality problem}
\begin{aligned}
    \max_{b,q}\quad&
       \overline\omega
       \int_{\underline\beta}^{b}\int_\Theta
          [\psi(\theta)v(q(\theta,\beta))
             -\widetilde\beta(\beta)q(\theta,\beta)]
          \,dF(\theta)\,dG(\beta)\\
    \text{subject to}\quad&
       b\in[\underline\beta,\overline\beta],\\
    &  q(\cdot,\beta)\ \text{nondecreasing},\qquad
       q(\cdot,\beta)\geq q^m(\cdot,\beta)
       \quad\text{for }\beta<b.
\end{aligned}
\end{equation}
Write $q^*$ for the optimizing quality schedule and $\beta^*$ for the largest maximizing cutoff in this relaxed problem.  Given $\Pi(\beta^*)=0$, the envelope conditions determine the corresponding prices. In the analysis, I will assume that these prices give the lowest consumer strictly positive utility whenever the firm operates.\footnote{A sufficient condition for this is
\begin{equation} \label{eq: sufficient ir}
    (\overline\beta-\underline\beta)
    \int_\Theta q^e(\theta,\underline\beta)\,dF(\theta)
    <
    \max_{q\geq0}
    \{\underline\theta v(q)-\overline\beta q\}.
\end{equation}
This condition bounds the firm's information rents relative to the surplus available from serving the lowest consumer. The appendix verifies that it ensures strictly positive utility for every consumer whenever the firm operates.}

\section{Optimal inequality-aware regulation}\label{sec: analysis firm}

My main theorem shows that optimal regulation takes the form of \emph{baseline regulation}. The regulator protects the basic options consumed by low-WTP consumers and leaves the firm to design its premium offers. The qualities of the basic options lie between the corresponding unregulated and efficient levels.

The required basic menu depends on the regulator's consumer-side redistributive preferences. A range of basic options may be needed when these preferences are weak, but a single protected option is sufficient when they are strong. The following definition formalizes the strong redistributive concerns sufficient for this simple form of regulation to be optimal.

\begin{definition}[Strong consumer-side redistributive concerns]
\label{def: strong redistribution}
Consumer-side redistributive concerns are \emph{strong} if there is $\theta_w\in(\underline\theta,\overline\theta)$ such that
\begin{enumerate}[(i)]
\item $\omega(\theta)=0$  for $\theta>\theta_w$, and \item $ t\longmapsto \frac{1}{F(t)} \int_{\underline\theta}^{t}\psi(\theta)\,dF(\theta)$ is weakly decreasing on $(\underline\theta,\theta_w].$
\end{enumerate}

\end{definition}

The first condition requires welfare weights to be concentrated on low-WTP consumers. The second requires that among them, expanding a group from the bottom of this range does not raise its average redistributive virtual value. For example, if the WTP density is weakly decreasing, the condition holds when equal weight is placed on a bottom population share of at most one half and zero weight on the remaining consumers.

\begin{theorem}[Baseline regulation]\label{thm: baseline regulation}
The following regulation is optimal:
\begin{enumerate}
\item For each $\beta<\beta^*$, the regulator requires only the basic options $(q^*(\theta,\beta),p^*(\theta,\beta))$ consumed by low-WTP types $\theta\leq\theta^*(\beta)$, where $\theta^*(\beta)<\overline\theta$ is given by \eqref{eq: baseline cutoff}. The firm freely designs its premium menu and supplies monopoly qualities $q^m(\theta,\beta)$ to consumers $\theta>\theta^*(\beta)$.

\item The basic qualities are between the unregulated and efficient levels: $ q^m(\theta,\beta)\leq q^*(\theta,\beta) \leq q^e(\theta,\beta). $ The basic qualities $q^*(\theta,\beta)$ are weakly decreasing in the firm's cost. \item Low-cost firms $\beta<\beta^*$ operate, and high-cost firms $\beta>\beta^*$ shut down.
\end{enumerate}
Under strong consumer-side redistributive concerns, requiring only the single option $(q^*(\underline\theta,\beta),p^*(\underline\theta,\beta))$ for each $\beta<\beta^*$ is sufficient.
\end{theorem}

Baseline regulation uses the firm's pricing incentives to support redistribution. A profit-maximizing firm seeks to extract surplus from high-WTP consumers; a regulator with decreasing consumer weights also places relatively little value on their rents. The surplus the firm extracts from premium buyers enables the regulator to require inexpensive, cross-subsidized basic options.

From an inequality-aware regulator's perspective, an unregulated market has two main shortcomings: the firm receives too much surplus  relative to consumers, and low-WTP consumers receive qualities that are too low. Baseline regulation addresses both. It improves the qualities available to low-WTP consumers and shifts surplus from the firm to consumers through the utility guarantee created by the basic options.

Indeed, conditional on operation, every consumer obtains strictly higher utility than under unregulated monopoly. The lowest type receives positive utility, and consumer incentive compatibility extends this gain to higher types because regulated qualities are no lower than monopoly qualities. Even consumers choosing premium options benefit: their ability to select a basic option limits what the firm can charge them.

The cutoff below which consumers choose basic options is
\begin{equation}\label{eq: baseline cutoff}
\theta^*(\beta)
=
\inf\left\{
t\in(\underline\theta,\overline\theta):
\ \phi(t)\geq0,\quad
\sup_{\underline\theta\leq s<t}
\frac{\int_s^t\psi(\theta)\,dF(\theta)}{F(t)-F(s)}
\leq
\frac{\widetilde\beta(\beta)}{\beta}\phi(t)
\right\}.
\end{equation}
Without pooling at the cutoff, this condition reduces to $\psi(\theta^*)=(\widetilde\beta/\beta)\phi(\theta^*)$: at this point, the regulator's preferred quality coincides with unregulated monopoly quality.

Baseline regulation also screens the firm's private cost. The regulator offers a choice of contracts with different basic option requirements. Lower-cost firms select contracts with higher basic qualities.

The appendix establishes both that the firm optimally supplies the proposed allocation under its assigned contract and that it prefers this contract to every other contract, even when it can optimally redesign its additional offers after a deviation. The key is that, holding actual cost fixed, contracts intended for lower-cost firms induce weakly greater total quality provision even after optimal menu redesign. Lower-cost firms therefore have a comparative advantage in these contracts. The regulator can then set the prices of the basic options to induce self-selection without restricting the firm's additional offers.

\paragraph{Effect of redistributive preferences.}
The basic qualities reflect two redistributive costs of quality provision. First, within the consumer side, raising basic quality increases the information rents that higher-WTP consumers must receive: their chosen options must give them at least the utility available from the basic options. This limits redistribution toward low-WTP consumers. Second, raising basic quality increases the information rents of lower-cost firms, which can produce that quality more cheaply. These rents shift surplus across the two market sides, from consumers to the firm. Both considerations favor limiting basic quality to improve the distribution of the resulting surplus.

These two redistributive costs appear in the marginal condition for basic quality, which applies outside pooling regions:
\begin{equation}\label{eq: baseline marginal condition}
    \psi(\theta)v'(q^*(\theta,\beta))
       =\widetilde\beta(\beta).
\end{equation}
Relative to the efficient condition $\theta v'(q^e)=\beta$, consumer-side redistributive preferences reduce the valuation of quality provision from $\theta$ to $\psi(\theta)\leq\theta$, while the redistributive cost of firm information rents raises its effective cost from $\beta$ to $\widetilde\beta(\beta)\geq\beta$. Both adjustments contribute to basic qualities being weakly below the efficient level.

The following proposition describes how these two redistributive concerns affect regulation. Say that consumer-side redistributive concerns \emph{become stronger} if the average consumer welfare weight is unchanged and the cumulative weight below every type weakly increases.\footnote{For nonincreasing weights $\omega_{\mathrm{old}}$ and $\omega_{\mathrm{new}}$ with the same mean, this requires $\int_{\underline\theta}^{t}\omega_{\mathrm{new}}(\theta)\,dF(\theta) \geq \int_{\underline\theta}^{t}\omega_{\mathrm{old}}(\theta)\,dF(\theta)$ for every $t\in\Theta$.}

\begin{proposition}[Effect of redistributive preferences]
\label{prop: stronger redistribution}
Suppose either
\begin{enumerate}[(i)]
\item consumer-side redistributive concerns become stronger, or \item the welfare weight on firm profit decreases.
\end{enumerate}
At every cost type operating under both policies, basic qualities weakly decrease and the mass of consumers choosing basic options weakly decreases. The lowest consumer's utility weakly increases, and the change in consumer utility is weakly decreasing in WTP. The operating cutoff $\beta^*$ also weakly decreases.
\end{proposition}

Perhaps counterintuitively, stronger consumer-side redistributive concerns lead to weakly lower basic quality. Higher-WTP consumers value quality more, so reducing basic quality makes these options less attractive to them and limits the rents they obtain from access to the basic menu. This gives the regulator more scope to require inexpensive basic options financed through revenues from premium options. For the lowest-WTP consumer, the price reduction at least compensates for the loss of quality. Lower basic quality can therefore support greater redistribution in utility.

Proposition~\ref{prop: stronger redistribution} also shows that the weight on firm profit affects redistribution within the consumer side. Favoring consumers more strongly relative to the firm can increase redistribution toward low-WTP consumers, even though relative consumer welfare weights are unchanged. Redistributive preferences across the two market sides therefore also shape how surplus is distributed among consumers.

These gains are conditional on operation. Inducing higher-cost firms to operate increases the information rents that must be left to lower-cost firms. A regulator that places less weight on firm profit may therefore make low-WTP consumers better off when the firm operates but increase the probability of shutdown.

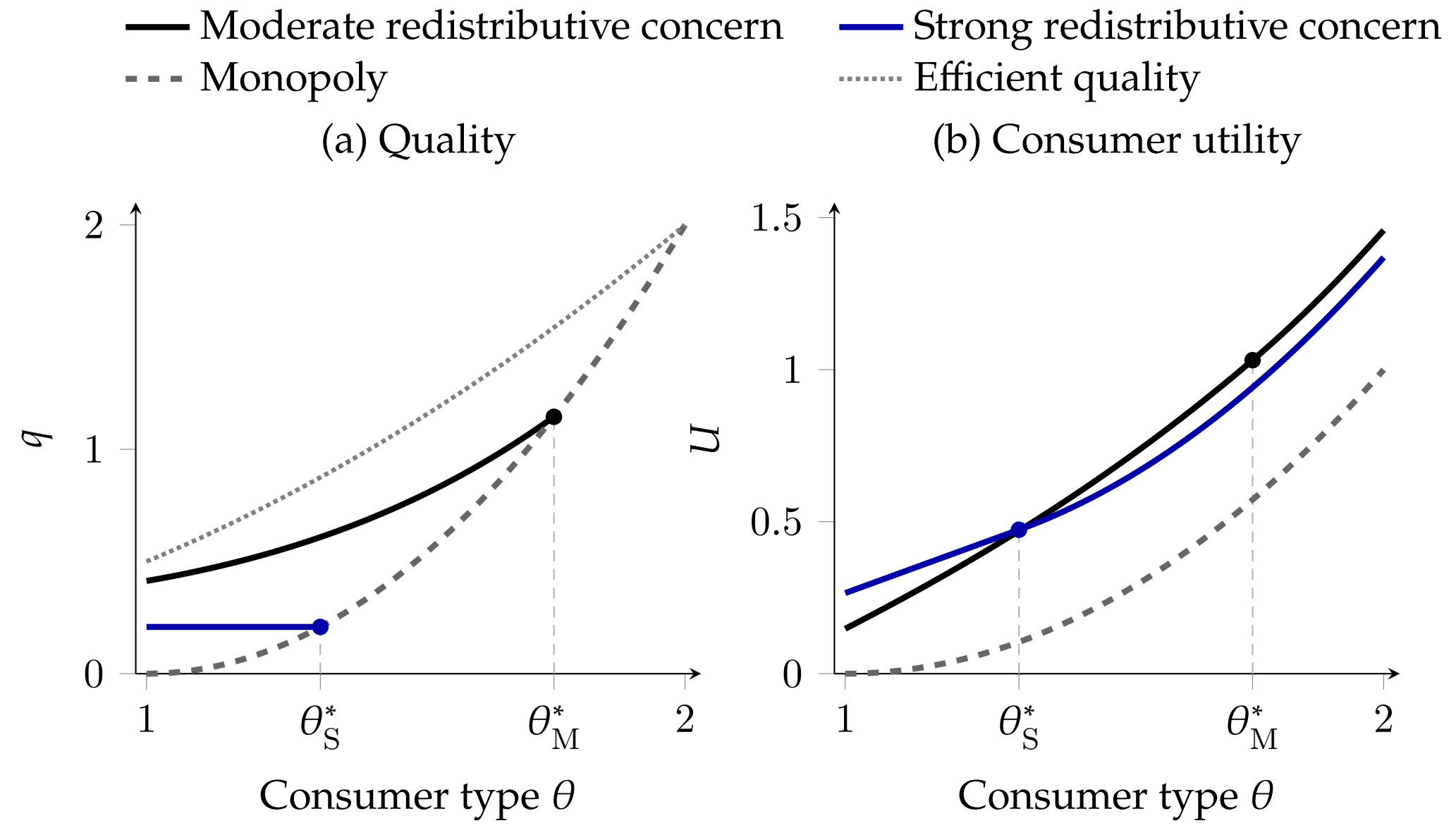
\begin{figure}[t]
\centering
    \begin{tikzpicture}
    \pgfmathsetmacro{\cutModerate}{(5.4-sqrt(3.56))/2}
    \pgfmathsetmacro{\cutStrong}{1+sqrt(5/48)}
    \pgfmathsetmacro{\utilityModerate}{0.147742721073871
        +((\cutModerate-1)+(\cutModerate-1)^2/4
        +(\cutModerate-1)^3/6)/1.1}
    \pgfmathsetmacro{\utilityStrong}{0.26517611560913745+5/24}
    \begin{groupplot}[
        group style={group size=2 by 1, horizontal sep=1.25cm},
        width=0.47\textwidth, height=6cm,
        axis lines=left,
        xlabel={Consumer type $\theta$},
        xmin=0.98, xmax=2.03,
        xtick={1,\cutStrong,\cutModerate,2},
        xticklabels={$1$,$\theta_{\mathrm S}^*$,
                     $\theta_{\mathrm M}^*$,$2$},
        grid=none, tick align=outside,
        samples=201,
        label style={font=\small},
        tick label style={font=\small},
        title style={font=\small},
        clip=false
    ]
    \nextgroupplot[
        title={(a) Quality}, ylabel={$q$},
        ymin=0, ymax=2.1, ytick={0,1,2},
        legend columns=2,
        legend style={at={(1.15,1.19)}, anchor=south,
            draw=none, fill=none, font=\small,
            /tikz/every even column/.append style={column sep=1em}},
        legend cell align=left
    ]
        \draw[black!30, thin, densely dashed]
            (axis cs:\cutStrong,0) -- (axis cs:\cutStrong,{5/24});
        \draw[black!30, thin, densely dashed]
            (axis cs:\cutModerate,0)
            -- (axis cs:\cutModerate,{2*(\cutModerate-1)^2});

        \addplot[black, ultra thick, domain=1:\cutModerate]
            {0.5*((0.5*x^2-0.5*x+1)/1.1)^2};
        \addlegendentry{Moderate redistributive concern}
        \addplot[blue!65!black, ultra thick, domain=1:\cutStrong]
            {5/24};
        \addlegendentry{Strong redistributive concern}
        \addplot[black!60, ultra thick, dashed, domain=1:2]
            {2*(x-1)^2};
        \addlegendentry{Monopoly}
        \addplot[black!50, very thick, densely dotted, domain=1:2]
            {x^2/2};
        \addlegendentry{Efficient quality}
        \addplot[only marks, mark=*, mark size=2pt, black,
            forget plot] coordinates
            {(\cutModerate,{2*(\cutModerate-1)^2})};
        \addplot[only marks, mark=*, mark size=2pt, blue!65!black,
            forget plot] coordinates {(\cutStrong,{5/24})};

    \nextgroupplot[
        title={(b) Consumer utility}, ylabel={$U$},
        ymin=0, ymax=1.55, ytick={0,0.5,1,1.5}
    ]
        \draw[black!30, thin, densely dashed]
            (axis cs:\cutStrong,0)
            -- (axis cs:\cutStrong,\utilityStrong);
        \draw[black!30, thin, densely dashed]
            (axis cs:\cutModerate,0)
            -- (axis cs:\cutModerate,\utilityModerate);

        \addplot[black, ultra thick, domain=1:2]
            {0.147742721073871
             +(min(x-1,\cutModerate-1)
               +min(x-1,\cutModerate-1)^2/4
               +min(x-1,\cutModerate-1)^3/6)/1.1
             +max((x-1)^2-(\cutModerate-1)^2,0)};
        \addplot[blue!65!black, ultra thick, domain=1:2]
            {0.26517611560913745
             +sqrt(5/12)*min(x-1,\cutStrong-1)
             +max((x-1)^2-(\cutStrong-1)^2,0)};
        \addplot[black!60, ultra thick, dashed, domain=1:2]
            {(x-1)^2};
        \addplot[only marks, mark=*, mark size=2pt, black,
            forget plot] coordinates {(\cutModerate,\utilityModerate)};
        \addplot[only marks, mark=*, mark size=2pt, blue!65!black,
            forget plot] coordinates {(\cutStrong,\utilityStrong)};
    \end{groupplot}
    \end{tikzpicture}
\caption{Baseline regulation under moderate and strong consumer-side redistributive concerns. The cutoffs $\theta_{\mathrm S}^*$ and $\theta_{\mathrm M}^*$ mark the highest types choosing basic options under strong and moderate redistributive concerns, respectively. Above each cutoff, the firm supplies monopoly qualities. Consumer WTP is uniform on $[1,2]$, firm cost is uniform on $[0.9,2.2]$, $v(q)=\sqrt{2q}$, and $\omega_F=0$; both panels fix $\beta=1$. Moderate concerns correspond to $\omega(\theta)=2.5-\theta$; strong concerns correspond to $\omega(\theta)=4\mathbf 1\{\theta\leq1.25\}$. Both weight functions have mean one.}
\label{fig: baseline quality}
\end{figure}

\paragraph{Single basic option.}
Consumer-side redistributive preferences also determine how extensive regulation needs to be. Under the strong concerns in Definition~\ref{def: strong redistribution}, Theorem~\ref{thm: baseline regulation} shows that requiring a single basic option is sufficient.

Both of the two conditions in the definition have a role for a single basic option being sufficient. Condition (ii) ensures that among low-WTP consumers, expanding a group from the bottom of the distribution does not raise its average redistributive virtual value. This favors concentrating quality provision toward the lowest types, whereas consumer incentive compatibility requires quality to be weakly increasing in WTP. Ironing resolves this: low-WTP consumers are pooled at a common basic option.

Condition (i) ensures that higher-WTP consumers receive zero welfare weight. The regulator therefore places no direct value on their rents and, like the firm, values the revenue they generate. Therefore,  no further required options are needed above the basic option. The resulting allocation is
\[
    q^*(\theta,\beta)
       =\max\{q^*(\underline\theta,\beta),q^m(\theta,\beta)\}.
\]
The firm supplies the premium qualities on its own, so a single required option suffices for each regulatory contract.

Figure~\ref{fig: baseline quality} illustrates this. With moderate redistributive concerns, the regulator protects a range of basic qualities. Under strong concerns, it requires one option chosen by a smaller group of low-WTP consumers. Consistent with Proposition~\ref{prop: stronger redistribution}, Panel~(a) shows that stronger redistributive concerns lower basic quality and contracts the  range of consumers choosing it. Panel~(b) shows that the accompanying price reduction more than compensates the lowest-WTP consumers for the loss of quality. In this example, higher-WTP consumers receive lower utility than under moderate redistribution, although they still benefit relative to monopoly.

\paragraph{Equal-weight benchmarks.}
The case of equal consumer weights further highlights how consumer-side redistributive preferences simplify optimal regulation. In the equal-weight case, $\psi(\theta)=\theta$. Optimal regulation can then never be implemented by requiring a single basic option from each operating firm type. Relative to strictly decreasing weights, equal weights call for more extensive regulation, requiring basic options for a broader range of consumer types and leaving less scope for the firm to design its premium offers.

If the firm also receives the same positive welfare weight, both redistributive adjustments disappear: $\psi=\theta$ and $\widetilde\beta=\beta$. The regulator then maximizes total surplus: efficient qualities are implemented and all cost types operate. The firm and high-WTP consumers receive information rents, but these rents carry no distributional cost when their recipients receive the same welfare weight as everyone else.

\paragraph{Private information and the scope of regulation.}
The firm's private information reduces the scope of optimal regulation. If costs were known, the regulator would determine the quality--price menu for all consumers and set prices so that the firm just breaks even, provided that  all consumers receive positive welfare weights. No premium range would be left for the firm to design on its own. With private costs, higher quality requirements mean  greater information rents for lower-cost firms or a higher probability of shutdown. Therefore, the regulator only requires the firm to offer basic options.

\section{Discussion}\label{sec: discussion}

\subsection{Fully flexible regulation mechanism} \label{sec: flexible}

In the main model, a regulatory contract specifies which quality--price options the firm must offer, but the firm may offer others. So, the option a consumer ultimately buys need not be one the regulatory contract specifies. This is a natural restriction: it is easier to verify that a product of a given quality is available at a specified price than to prevent the firm and a consumer from agreeing on terms they both prefer. I now ask what the regulator gains from full control over the terms of trade.

Under fully flexible regulation, the regulator can commit to any mechanism governing trade between the firm and consumers, and the parties cannot contract on other terms. I consider Bayes--Nash equilibrium outcomes with voluntary participation on both sides of the market.\footnote{As in the main analysis, condition~\eqref{eq: sufficient ir} ensures that consumer participation is slack.} By the revelation principle, it suffices to consider direct mechanisms with simultaneous reports of the firm's cost and consumers' WTP in which truthful reporting is an equilibrium. For each profile of reports, the mechanism specifies a joint distribution over the qualities the firm supplies and the prices consumers pay the firm.

I show in the appendix that even in this class, an optimal mechanism can be implemented by offering the firm a choice among deterministic menus of quality--price options. An operating firm must offer exactly the menu it selects. Each consumer then chooses an option from that menu or opts out.

The ability to prohibit side contracting allows the regulator to use premium quality to screen the firm. Consider a firm with cost $\beta$ that selects the menu intended for type $\hat\beta>\beta$. It collects the same payments as type $\hat\beta$ but saves $\hat\beta-\beta$ on each unit of quality it supplies. Its profit on this menu therefore exceeds that of the intended type by $(\hat\beta-\beta)\int_\Theta q(\theta,\hat\beta)\,dF(\theta)$. The more quality this menu assigns to high-WTP consumers, the more a low-cost firm saves by choosing it. Restricting premium quality under the high-cost contract therefore reduces the information rents needed to induce truthful reporting. In the main model, however, the regulator cannot decrease premium quality of the high-cost firm below the monopoly level: the firm would find it profitable to supplement the required menu with additional offers.

\begin{proposition}[Fully flexible regulation]
\label{prop: fully flexible regulation}
Under fully flexible regulation, the optimal quality allocation solves \eqref{eq: relaxed quality problem} without the constraint $q\geq q^m$. All cost types operate. At each cost $\beta$, consumers with $\theta>\theta^*(\beta)$ receive strictly lower quality than under unregulated monopoly, where $\theta^*(\beta)$ is given by \eqref{eq: baseline cutoff}.
\end{proposition}

The reason for restricting premium quality is clearest at the top of the WTP distribution. An unregulated monopolist supplies efficient quality to the highest-WTP consumer. From the regulator's perspective, however, supplying more quality under a given cost report makes that report more attractive to lower-cost firms. The regulator therefore evaluates top quality at the virtual cost rather than the actual cost:
\[
    q(\overline\theta,\beta)
    =(v')^{-1}\!\left(\frac{\widetilde\beta(\beta)}{\overline\theta}\right)
    <(v')^{-1}\!\left(\frac{\beta}{\overline\theta}\right)
    =q^m(\overline\theta,\beta).
\]
So, the mechanism places a cap on quality for each cost type: the highest quality the firm may sell is strictly below what it would offer its highest-WTP consumer under unregulated monopoly. The comparison extends beyond the top consumer $\overline \theta$, to every consumer whose option the firm designs freely under baseline regulation (Theorem~\ref{thm: baseline regulation}).

The ability to restrict premium quality also changes the shutdown decision. Under baseline regulation, keeping a high-cost firm in the market entails allowing premium sales that make its contract attractive to lower-cost firms. The regulator may prefer shutdown to leaving those firms the resulting information rents. Under full control, it can instead limit the premium sales of a high-cost contract. Providing a modest level of quality is then preferable to shutdown even at high costs.

\subsection{Greater weight on high-WTP consumers}
\label{sec: high wtp weights}

The main analysis assumes that consumer welfare weights weakly decrease with WTP. The regulator may instead favor high-WTP consumers, for example when WTP reflects need rather than ability to pay. This changes the case for leaving premium options to the firm. Under baseline regulation, the firm's incentive to extract surplus from premium buyers allows the regulator to require low prices for basic options. Here, the regulator would instead like those buyers to receive more surplus.

\begin{remark}[Premium regulation]
\label{rem: premium regulation}
Suppose the firm's cost is known, consumer participation is slack, and there are strong redistributive concerns for high-WTP consumers: there exists $\theta_w\in(\underline\theta,\overline\theta)$ such that $\omega(\theta)=0$ for $\theta<\theta_w$ and $\psi$ is strictly decreasing on $[\theta_w,\overline\theta]$.\footnote{For example, if all consumers above $\theta_w$ receive the same positive weight, this condition holds for $\theta_w$ sufficiently close to $\overline\theta$.} It is optimal to require a single premium option, sold below cost and chosen by high-WTP consumers $\theta\geq\theta_P$ for some $\theta_P\in(\underline\theta,\theta_w)$, while leaving the firm free to design and price its lower-quality options. The required premium quality exceeds even the efficient quality of the highest-WTP consumer.
\end{remark}

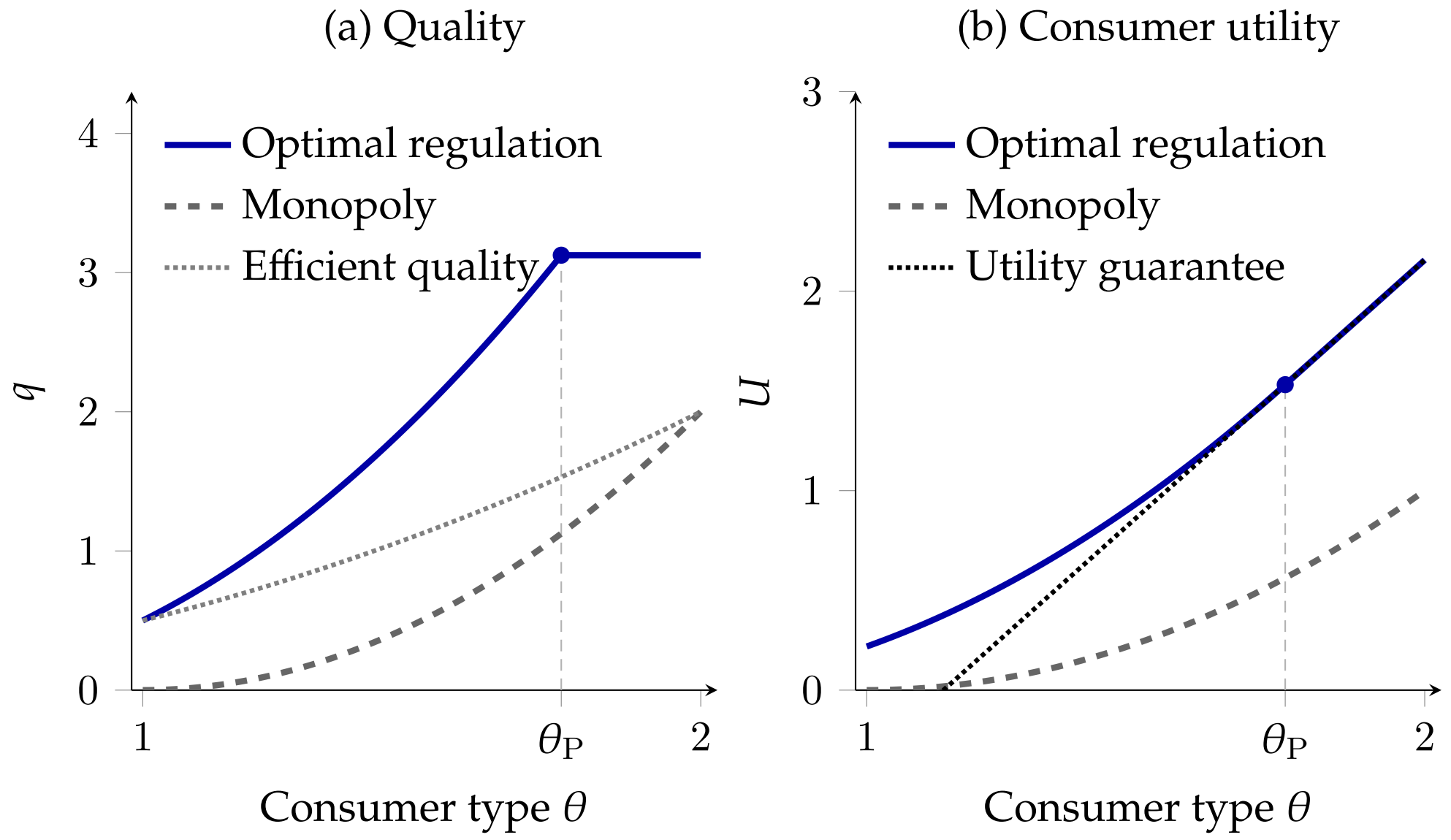
\begin{figure}[t]
\centering
    \begin{tikzpicture}
    \pgfmathsetmacro{\premiumCutoff}{7/4}
    \pgfmathsetmacro{\premiumQuality}{25/8}
    \pgfmathsetmacro{\premiumUtility}{49/32}
    \begin{groupplot}[
        group style={group size=2 by 1, horizontal sep=1.25cm},
        width=0.47\textwidth, height=7cm,
        axis lines=left,
        xlabel={Consumer type $\theta$},
        xmin=0.98, xmax=2.03,
        xtick={1,\premiumCutoff,2},
        xticklabels={$1$,$\theta_{\mathrm P}$,$2$},
        grid=none, tick align=outside,
        samples=201,
        label style={font=\small},
        tick label style={font=\small},
        title style={font=\small},
        legend style={at={(0.03,0.97)}, anchor=north west,
            draw=none, fill=none, font=\small, row sep=2pt},
        legend cell align=left,
        clip=true
    ]
    \nextgroupplot[
        title={(a) Quality}, ylabel={$q$},
        ymin=0, ymax=4.3, ytick={0,1,2,3,4}
    ]
        \draw[black!30, thin, densely dashed]
            (axis cs:\premiumCutoff,0)
            -- (axis cs:\premiumCutoff,\premiumQuality);
        \addplot[blue!65!black, ultra thick,
            domain=1:\premiumCutoff] {(2*x-1)^2/2};
        \addlegendentry{Optimal regulation}
        \addplot[blue!65!black, ultra thick, forget plot,
            domain=\premiumCutoff:2] {25/8};
        \addplot[black!60, ultra thick, dashed, domain=1:2]
            {2*(x-1)^2};
        \addlegendentry{Monopoly}
        \addplot[black!50, very thick, densely dotted, domain=1:2]
            {x^2/2};
        \addlegendentry{Efficient quality}
        \addplot[blue!65!black, only marks, mark=*, mark size=2pt,
            forget plot] coordinates
            {(\premiumCutoff,\premiumQuality)};

    \nextgroupplot[
        title={(b) Consumer utility}, ylabel={$U$},
        ymin=0, ymax=3.0, ytick={0,1,2,3}
    ]
        \draw[black!30, thin, densely dashed]
            (axis cs:\premiumCutoff,0)
            -- (axis cs:\premiumCutoff,\premiumUtility);
        \addplot[blue!65!black, ultra thick,
            domain=1:\premiumCutoff] {7/32+x^2-x};
        \addlegendentry{Optimal regulation}
        \addplot[blue!65!black, ultra thick, forget plot,
            domain=\premiumCutoff:2] {5*x/2-91/32};
        \addplot[black!60, ultra thick, dashed, domain=1:2]
            {(x-1)^2};
        \addlegendentry{Monopoly}
        \addplot[black, very thick, densely dotted, domain=1:2]
            {5*x/2-91/32};
        \addlegendentry{Utility guarantee}
        \addplot[blue!65!black, only marks, mark=*, mark size=2pt,
            forget plot] coordinates
            {(\premiumCutoff,\premiumUtility)};
    \end{groupplot}
    \end{tikzpicture}
\caption{Premium regulation. Consumers with $\theta\geq\theta_{\mathrm P}$ choose the required option; the firm designs the lower-quality offers. Panel~(a) compares qualities. In panel~(b), the dotted line gives utility from the required option and coincides with regulated utility above $\theta_{\mathrm P}$.  Consumer WTP is uniform on $[1,2]$, the firm's cost is known to be $\beta=1$, $v(q)=\sqrt{2q}$, $\omega_F=0$, and $\omega(\theta)=8\mathbf 1\{\theta\geq15/8\}$.}
\label{fig: premium regulation}
\end{figure}

To understand the remark, first allow the regulator to fully control the terms of trade, as in Section~\ref{sec: flexible}. With known costs, the regulator leaves the firm zero profit. By the same ironing logic as under strong consumer-side redistributive concerns in Theorem~\ref{thm: baseline regulation}, consumers with positive welfare weight pool on a common premium option.

Requiring this premium option alone implements the full-control optimum. Every consumer with positive welfare weight necessarily obtains at least the utility from the required option. The firm can offer the same menu as under full control and break even. If a different menu gave it a positive profit, the regulator under full control could adopt that menu and return the profit to consumers through lower prices, contradicting optimality.

Figure~\ref{fig: premium regulation} illustrates the policy. In panel~(a), high-WTP consumers choose the same premium quality, above the efficient level, while lower types choose the options designed by the firm. Panel~(b) compares consumer utilities under monopoly and optimal regulation and shows the linear utility guarantee from the required option.  All consumers benefit relative to monopoly, but high-WTP consumers benefit more.

The assumption of known costs matters: with private costs, lowering prices under one contract can change the firm's incentive to select it. I leave a full characterization of optimal regulation under increasing welfare weights for future research.

\begin{appendix}
\renewcommand*\appendixpagename{\Large Appendix}\label{sec:appendixa}
\appendixpage
\noappendicestocpagenum
\addappheadtotoc

\section{Proofs}

\startcontents[appendix]
\printcontents[appendix]{l}{1}{\setcounter{tocdepth}{2}}

\subsection{Proof of Theorem~\ref{thm: baseline regulation}}
\label{app: baseline regulation}

\begin{proof}
The preliminary analysis shows that the solution to the optimization problem \eqref{eq: relaxed quality problem} gives an upper bound on the regulator's payoff. I first characterize that solution and then show that the required options in the theorem implement it (even when the firm can change both its report and its additional offers).

\noindent\paragraph{Optimal qualities and the baseline cutoff $\theta^*$.}
Fix a cost type $\beta$. To characterize quality provision, consider which consumers receive at least a given quality $z>0$. Since quality must be nondecreasing in WTP, these consumers form an upper interval. Under monopoly this interval starts at $\phi^{-1}(\beta/v'(z))$. (Here, $\phi^{-1}$ is taken to equal $\underline\theta$ or $\overline\theta$ when its argument lies below or above the range of $\phi$.)

Because regulation cannot lower quality below monopoly provision, the regulated cutoff cannot exceed $\phi^{-1}(\beta/v'(z))$. The regulator's marginal payoff from increasing quality at $z$ for type $\theta$ is $\psi(\theta)v'(z)-\widetilde\beta(\beta)$. For each quality level $z$, the optimal cutoff in consumer type is therefore the largest maximizer
\begin{equation}\label{eq: appendix quality thresholds}
 a(z,\beta)\in
 \arg\max_{\underline\theta\leq a\leq
                     \phi^{-1}(\beta/v'(z))}
 \int_a^{\overline\theta}
 [\psi(\theta)v'(z)-\widetilde\beta(\beta)]\,dF(\theta).
\end{equation}
The original objective is the integral of these marginal payoffs over all quality levels. These choices therefore solve the allocation problem, provided the cutoffs are nondecreasing in $z$, as verified next.

Divide the objective in \eqref{eq: appendix quality thresholds} by $v'(z)>0$. As $z$ increases, the term $\widetilde\beta(\beta)/v'(z)$ subtracted from $\psi$ increases. This makes a higher cutoff more attractive relative to a lower one. The upper bound on the cutoff also increases, so the largest maximizing cutoff cannot decrease. The same argument applies when $\beta$ increases. Hence $a(z,\beta)$ is nondecreasing in both arguments, and $q^*(\theta,\beta)$ is nonincreasing in $\beta$. Strict convexity of $v^{-1}$ makes the optimal allocation unique almost everywhere.

The allocation cannot exceed $q^e$. Replacing it by $\min\{q^*,q^e\}$ preserves monotonicity and the monopoly lower bound, and increases the objective wherever $q^*>q^e$, since $\psi\leq\theta$ and $\widetilde\beta(\beta)\geq\beta$. Hence $q^m\leq q^*\leq q^e$. The allocation is also continuous in the interior. Otherwise, all quality levels between the two sides of a jump would have the same interior cutoff in \eqref{eq: appendix quality thresholds}. Its first-order condition, $\psi(a)=\widetilde\beta(\beta)/v'(z)$, cannot hold at the same $a$ for an interval of distinct $z$.

I next show that regulation raises quality above the monopoly level only for a lower interval of consumer types. Two monotonicity properties are useful. First, $\psi-\phi$ is nonincreasing, because
\[
 \psi(\theta)-\phi(\theta)
 =
 \frac{\displaystyle
       \int_\theta^{\overline\theta}\omega(t)\,dF(t)}
      {\overline\omega[1-F(\theta)]}
 \frac{1-F(\theta)}{f(\theta)}.
\]
The first factor is the average welfare weight among types above $\theta$, divided by $\overline\omega$; it decreases with $\theta$. The second factor decreases by log-concavity of $f$. Second, $\widetilde\beta(\beta)-\beta$ is nondecreasing by log-concavity of $g$.

Fix a quality $z$ whose monopoly cutoff $t=\phi^{-1}(\beta/v'(z))$ exceeds $\underline\theta$. Regulation leaves this cutoff unchanged if and only if
\begin{equation}\label{eq: appendix unchanged provision}
 \int_s^t
 \left[\psi(\theta)
       -\frac{\widetilde\beta(\beta)}{v'(z)}\right]dF(\theta)
 \leq0
 \quad\text{for every }s\in[\underline\theta,t).
\end{equation}
Suppose this holds and $t<\overline\theta$. Dividing by $F(t)-F(s)$ and letting $s$ approach $t$ gives $\psi(t)\leq\widetilde\beta(\beta)/v'(z)$. Since $\phi(t)=\beta/v'(z)$, it follows that
\[
 \psi(t)-\phi(t)
 \leq\frac{\widetilde\beta(\beta)-\beta}{v'(z)}.
\]
Now consider a higher quality $z'>z$. Replacing $z$ by $z'$ lowers the integrand in \eqref{eq: appendix unchanged provision} at every consumer type. For consumers $\theta$ between $t$ and the new monopoly cutoff, the two monotonicity properties imply
\[
 \psi(\theta)-\frac{\widetilde\beta(\beta)}{v'(z')}
 \leq \psi(t)-\phi(t)
       -\frac{\widetilde\beta(\beta)-\beta}{v'(z')}
 \leq0.
\]
So,  \eqref{eq: appendix unchanged provision} also holds at $z'$. The same reasoning applies if cost increases instead of quality: both $\widetilde\beta$ and $\widetilde\beta-\beta$ increase. If $t=\overline\theta$, there is no additional interval to consider.

Consequently, if the regulated cutoff equals the monopoly cutoff at some quality level above the lowest monopoly quality, the cutoffs are also equal at every higher quality level. There is therefore a consumer cutoff above which $q^*=q^m$ and below which $q^*>q^m$, except possibly where both qualities are zero. The highest required quality is nonincreasing in cost; this also accounts for the quality already supplied to every consumer under monopoly, which itself decreases with cost. Substituting $1/v'(z)=\phi(t)/\beta$ into \eqref{eq: appendix unchanged provision} gives the formula for $\theta^*(\beta)$ in \eqref{eq: baseline cutoff}. The case of zero quality follows by taking $z\downarrow0$.

This cutoff is strictly below $\overline\theta$. Indeed, if $1/v'(z)>\overline\theta/\widetilde\beta(\beta)$, the integrand in \eqref{eq: appendix quality thresholds} is negative for every consumer type, so regulation leaves provision at that quality level to the firm. Since $\widetilde\beta(\beta)>\beta$ for every cost type, this applies to an interval of qualities strictly below the top monopoly quality. Continuity of $q^*$ also shows that the required qualities form an interval.

\medskip
\noindent\paragraph{Required options and prices.}
Take the firm's profit $\Pi$ from \eqref{eq: minimum firm rent}, using $q=q^*$ and setting quality to zero above $\beta^*$. The corresponding bottom utility is
\begin{equation}\label{eq: appendix bottom utility}
 U_0(\beta)=
 \int_\Theta
 [\phi(\theta)v(q^*(\theta,\beta))
       -\beta q^*(\theta,\beta)]\,dF(\theta)-\Pi(\beta).
\end{equation}
Equation~\eqref{eq: setup consumer envelope} then determines $U$, and prices are $p^*(\theta,\beta)=\theta v(q^*(\theta,\beta))-U(\theta,\beta)$. These qualities and prices satisfy consumer incentive compatibility. The maintained participation condition ensures $U_0(\beta)>0$.\footnote{As noted in the text, it is sufficient that $ (\overline\beta-\underline\beta) \int_\Theta q^e(\theta,\underline\beta)\,dF(\theta) <\max_{q\geq0} \{\underline\theta v(q)-\overline\beta q\}$. To see this, the virtual-profit integrand decreases in quality above $q^m$. The quality bounds therefore make the integral in \eqref{eq: appendix bottom utility} at least $\int_\Theta [\phi(\theta)v(q^e(\theta,\beta)) -\beta q^e(\theta,\beta)]\,dF(\theta) = \max_{q\geq0}\{\underline\theta v(q)-\beta q\}$. Furthermore, \eqref{eq: minimum firm rent} bounds $\Pi(\beta)$ by the left-hand side of the stated condition.}

For each report $\hat\beta<\beta^*$, require only the options assigned to $\theta\leq\theta^*(\hat\beta)$. Their guarantee $\underline U(\theta,\hat\beta)$ equals $U(\theta,\hat\beta)$ below the cutoff. Above it, the guarantee is the utility from choosing the last required option. Hence
\begin{equation}\label{eq: appendix guarantee slopes}
 \underline U_\theta(\theta,\hat\beta)
 =
 v\!\left(
   \min\{q^*(\theta,\hat\beta),
          q^*(\theta^*(\hat\beta),\hat\beta)\}
   \right).
\end{equation}
Both qualities in this minimum are nonincreasing in the report, as established above.

\noindent\paragraph{The firm's choice of additional options.}
Fix $\beta$ and suppose the firm chooses its assigned contract. I show that its optimal response is the proposed allocation, so that $\pi(\beta;\beta)=\Pi(\beta)$. For this step, define
\[
 y(\theta)=f(\theta)
   \left[\frac{\beta}{v'(q^*(\theta,\beta))}
                       -\phi(\theta)\right],
 \qquad 1/v'(0):=0.
\]
The expression in brackets is the marginal loss in virtual profit from increasing the utility slope. The quality bounds imply $0\leq y\leq1-F$. I next verify that $y$ is nonincreasing.

Where consumers receive different qualities strictly above monopoly provision, the marginal condition from \eqref{eq: appendix quality thresholds} gives $\psi(\theta)v'(q^*(\theta,\beta)) =\widetilde\beta(\beta)$. Differentiating the resulting expression for $y$ yields
\begin{align*}
 y'(\theta)
 =-f(\theta)\Bigg[
 \frac{\beta}{\widetilde\beta(\beta)}
              \frac{\omega(\theta)}{\overline\omega}+\left(1-\frac{\beta}{\widetilde\beta(\beta)}\right)
            \left(2+\theta\frac{f'(\theta)}{f(\theta)}\right)
              \Bigg]
 \leq
 -\left(1-\frac{\beta}{\widetilde\beta(\beta)}\right)f(\theta)
 \leq0.
\end{align*}
The first inequality follows from decreasing welfare weights, the increasing hazard rate, and $y\geq0$.\footnote{Use $\omega(\theta)\geq \int_\theta^{\overline\theta}\omega(t)\,dF(t)/[1-F(\theta)]$ and $f'/f\geq-f/(1-F)$. Together with $y\geq0$, these imply $\frac{\beta}{\widetilde\beta(\beta)} \frac{\omega(\theta)}{\overline\omega} \geq (1-\frac{\beta}{\widetilde\beta(\beta)}) \frac{f(\theta)\phi(\theta)}{1-F(\theta)}$. Substitution gives the displayed bound.} At points where quality is positive and equal to monopoly level, $y=0$. On an interval of constant quality, differentiation instead gives
\[
 y'=f'\left[\frac{\beta}{v'(q^*)}-\theta\right]-2f\leq0.
\]
If $f'\geq0$, this follows from $q^*\leq q^e$. If $f'<0$, use $q^*\geq q^m$ and $f'/f\geq-f/(1-F)$. Outside constant-quality intervals, $y$ is the positive part of $f[(\beta/\widetilde\beta(\beta))\psi-\phi]$. The same derivative calculation makes this positive part nonincreasing. Continuity of the allocation ensures that $y$ is continuous at the boundaries of the constant-quality intervals. Hence $y$ is nonincreasing throughout.

For $\theta>\theta^*(\beta)$, the firm supplies monopoly quality, so $y=0$. So $y$ can decrease only where $U=\underline U$, and $y(\overline\theta)=0$. Consider any alternative feasible consumer utility $u\geq\underline U(\cdot,\beta)$. Concavity of the profit expression in \eqref{eq: firm continuation value}, and integration by parts, gives
\begin{align*}
 &\left\{
  \int_\Theta[\phi u'-\beta v^{-1}(u')]\,dF
         -u(\underline\theta)
  \right\}-\Pi(\beta)\\
 &\qquad\leq
 [y(\underline\theta)-1]
       [u(\underline\theta)-U(\underline\theta,\beta)]
 +\int_{\underline\theta}^{\overline\theta}
       [u(\theta)-U(\theta,\beta)]\,dy(\theta)
 \leq0.
\end{align*}
The final inequality uses $y(\underline\theta)\leq1$, $dy\leq0$, and equality of $U$ and its guarantee wherever $y$ decreases. No additional offers can therefore raise the firm's profit under its assigned contract.

\noindent\paragraph{Contract choice and shutdown.}
It remains to allow the firm to choose a different contract and then redesign its additional offers. Equation~\eqref{eq: appendix guarantee slopes} and Lemma~\ref{lem: ordered guarantees} below imply that, holding actual cost fixed, a lower report induces weakly greater quality provision for every consumer. By the cost envelope theorem, the derivative of $\pi(\hat\beta;s)$ with respect to actual cost $s$ is minus total quality provision after the firm optimizes its additional offers. Consequently, for $s>\hat\beta$ this derivative is weakly below $\Pi'(s)$; for $s<\hat\beta$ it is weakly above. Using $\pi(\hat\beta;\hat\beta)=\Pi(\hat\beta)$ gives
\begin{equation}\label{eq: appendix global IC}
 \Pi(\beta)-\pi(\hat\beta;\beta)
 =
 \int_{\hat\beta}^{\beta}
 \left[\Pi'(s)
       -\frac{\partial\pi(\hat\beta;s)}{\partial s}\right]ds
 \geq0.
\end{equation}
If $\beta>\hat\beta$, the integrand is nonnegative. If $\beta<\hat\beta$, it is nonpositive and the limits of integration are reversed. This verifies firm incentive compatibility for every report.

Operating types earn nonnegative profit. Taking $\beta\uparrow\beta^*$ in \eqref{eq: appendix global IC} gives $\pi(\hat\beta;\beta^*)\leq0$ for every offered contract. Since profit under any contract decreases with actual cost, types above $\beta^*$ cannot gain from operating. The mechanism therefore implements the allocation and operating cutoff from \eqref{eq: relaxed quality problem}, and attains its welfare bound.

\noindent\paragraph{A single required option.}
Finally, suppose consumer-side redistributive concerns are strong. For any fixed quality $z$, maximizing \eqref{eq: appendix quality thresholds} is equivalent to minimizing
\[
 \int_{\underline\theta}^a
 \left[\psi(\theta)-\frac{\widetilde\beta(\beta)}{v'(z)}\right]
 \,dF(\theta)
 =
 F(a)\left[
   \frac{1}{F(a)}
       \int_{\underline\theta}^a\psi(\theta)\,dF(\theta)
   -\frac{\widetilde\beta(\beta)}{v'(z)}
       \right].
\]
The feasible interval is unchanged, and the expression is zero at $a=\underline\theta$. For $a\leq\theta_w$, the definition of strong concerns makes the average inside the brackets nonincreasing in $a$. If the expression is negative, increasing $a$ makes the bracketed term weakly smaller and multiplies it by a larger population mass. If it is nonnegative everywhere, its minimum is zero at $\underline\theta$. Thus the minimum over any interval beginning at $\underline\theta$ and ending at or below $\theta_w$ is attained at one of its endpoints.

Above $\theta_w$, welfare weights are zero and $\psi=\phi$. Up to $\phi^{-1}(\beta/v'(z))$, the integrand in the expression being minimized is then nonpositive, so the integral is nonincreasing in $a$. Its largest minimizer is therefore either $\underline\theta$ or the monopoly cutoff. Every quality increase is supplied either to all consumers or only to those who receive it under monopoly. Hence
\[
 q^*(\theta,\beta)
 =
 \max\{q^*(\underline\theta,\beta),q^m(\theta,\beta)\}.
\]
Consumers with $\theta\leq\theta^*(\beta)$ receive a common quality and, by consumer incentive compatibility, a common price. The baseline menu therefore consists of a single option. The implementation and incentive-compatibility arguments above apply without change.
\end{proof}

\medskip \noindent The following lemma establishes the comparison between contracts used in the proof.

\begin{lemma}\label{lem: ordered guarantees}
Fix the firm's actual cost $\beta$ and consider two contracts $\hat\beta_1,\hat\beta_2$ whose required qualities are bounded and whose utility guarantees are strictly positive at $\underline\theta$. If
\[
 \underline U_\theta(\theta,\hat\beta_1)
 \geq \underline U_\theta(\theta,\hat\beta_2)
 \quad\text{almost everywhere},
\]
then, conditional on operating under each contract, the firm optimally supplies weakly higher quality to almost every consumer under contract $\hat\beta_1$.
\end{lemma}

\begin{proof}
First omit convexity of the firm's chosen utility $u$, but keep $u'\geq0$ and $u\geq\underline U(\cdot,\hat\beta)$. I show that the solution is nevertheless convex and then compare the solutions under the two contracts.

The firm's profit has two equivalent expressions:
\begin{align*}
 &\int_\Theta
   [\phi(\theta)u'(\theta)
      -\beta v^{-1}(u'(\theta))]\,dF(\theta)
      -u(\underline\theta)\\
 &\qquad=
 \int_\Theta
   \left[\left(\theta+\frac{F(\theta)}{f(\theta)}\right)
        u'(\theta)-\beta v^{-1}(u'(\theta))\right]\,dF(\theta)
      -u(\overline\theta).
\end{align*}
The equality follows from $u(\overline\theta)-u(\underline\theta)=\int_\Theta u'(\theta)\,d\theta$. These expressions give the quality bounds
\begin{equation}\label{eq: ordered guarantees quality bounds}
 \phi(\theta)
 \leq\frac{\beta}{v'(q(\theta))}
 \leq\theta+\frac{F(\theta)}{f(\theta)},
 \qquad 1/v'(0):=0.
\end{equation}
For the first inequality, raise any quality below its monopoly level while holding $u(\underline\theta)$ fixed. This raises consumer utilities and increases the first profit expression. For the second, reduce any quality violating the upper bound while holding $u(\overline\theta)$ fixed. This also raises consumer utilities and increases the second profit expression. Both adjustments therefore preserve the utility guarantee $u\geq\underline U$.

The upper bound is finite. These adjustments therefore allow a sufficiently large uniform quality bound to be imposed without changing the problem. With qualities bounded, profit tends to minus infinity as $u(\underline\theta)$ increases. Since a constant utility satisfying the guarantee gives finite profit, utility levels can also be bounded without loss. Compactness then gives an optimum. Strict convexity of $v^{-1}$ makes the optimal slopes $u'$ unique almost everywhere. Given these slopes, the firm chooses the lowest feasible utility level, so the optimal utility curve is unique.

Consider a closed interval where the utility guarantee $u\geq\underline U$ is strictly slack. Small changes in utility preserve this inequality. Holding utility fixed at the interval's endpoints fixes the integral of $u'$. The usual first-order conditions therefore give a constant $\lambda$ such that quality maximizes
\[
 \left[\phi(\theta)+\frac{\lambda}{f(\theta)}\right]v(q)
 -\beta q
\]
at almost every type in the interval. If quality is positive somewhere, \eqref{eq: ordered guarantees quality bounds} implies $0\leq\lambda\leq1$. Hence
\[
 \phi+\frac{\lambda}{f}
 =(1-\lambda)\phi
   +\lambda\left(\theta+\frac{F}{f}\right)
\]
is nondecreasing by log-concavity. Quality, and therefore $u'$, is nondecreasing on the interval. If quality is zero throughout, the same conclusion is immediate.

So, $u$ is convex wherever it lies strictly above the guarantee. Since the guarantee $\underline U$ itself is convex, $u$ is convex everywhere. Omitting consumer incentive compatibility has consequently not changed the solution.

Let $u_1,u_2$ be the optimal utilities under the two contracts. I next show that $u_1-u_2$ is nondecreasing. Suppose otherwise. There would be consumer types $a<t<b$ and a constant $c$ such that
\[
 u_1(a)>u_2(a)+c,\qquad
 u_1(t)=u_2(t)+c,\qquad
 u_1(b)<u_2(b)+c.
\]
The hypothesis implies that $\underline U(\theta,\hat\beta_1) -\underline U(\theta,\hat\beta_2)$ is nondecreasing.

Suppose first that $\underline U(t,\hat\beta_1) \leq\underline U(t,\hat\beta_2)+c$. The same inequality holds for every $\theta\leq t$. On $[\underline\theta,t]$, replace the two utility curves by
\[
 \min\{u_1,u_2+c\},
 \qquad
 \max\{u_1-c,u_2\},
\]
respectively, leaving them unchanged above $t$. Both replacements remain nondecreasing, satisfy their respective guarantees, and match the original utilities at $t$. They are therefore feasible in the relaxed problems. Their utility levels sum to the original sum at every type, and their slopes are either unchanged or exchanged. The sum of the two profits is consequently unchanged. Since both original choices were optimal, both replacements must also be optimal. Uniqueness then contradicts the strict change in the first utility at $a$.

If instead $\underline U(t,\hat\beta_1) >\underline U(t,\hat\beta_2)+c$, this inequality holds for every $\theta\geq t$. Apply the same argument on $[t,\overline\theta]$, replacing the utilities there by
\[
 \max\{u_1,u_2+c\},
 \qquad
 \min\{u_1-c,u_2\}.
\]
Both remain feasible and the sum of profits is unchanged, now contradicting the strict change at $b$.

Hence $u_1-u_2$ is nondecreasing, so $u_1'\geq u_2'$ almost everywhere. Since $u_i'=v(q_i)$ and $v$ is increasing, the stated quality comparison follows.
\end{proof}

\subsection{Proof of Proposition~\ref{prop: stronger redistribution}}
\label{app: stronger redistribution}

\begin{proof}
Stronger consumer-side redistributive concerns weakly reduce the total welfare weight above every consumer type while keeping $\overline\omega$ fixed. They therefore weakly lower $\psi$ and leave $\widetilde\beta$ unchanged. A decrease in $\omega_F$ instead weakly raises $\widetilde\beta$ and leaves $\psi$ unchanged. Neither change affects the monopoly allocation.

First compare qualities conditional on operation. For each quality level $z$, either change weakly lowers the integrand in \eqref{eq: appendix quality thresholds}, while leaving the feasible interval unchanged. The difference between the new and old expressions being maximized is therefore nondecreasing in their lower limit. Hence the largest maximizing type $a(z,\beta)$ cannot decrease. The set of consumers receiving quality above $z$ can only shrink, so $q^*(\theta,\beta)$ weakly decreases. This comparison also applies when consumers are pooled.

In \eqref{eq: baseline cutoff}, the averages on the left-hand side weakly decrease, while the right-hand side weakly increases for every $t$ with $\phi(t)\geq0$. Every type satisfying the cutoff condition before the change therefore continues to satisfy it afterward. Consequently, $\theta^*(\beta)$ weakly decreases, as does the mass $F(\theta^*(\beta))$ of consumers choosing basic options.

Next consider the operating cutoff. At each cost, either change weakly reduces the contribution of every feasible quality allocation to \eqref{eq: relaxed quality problem}, so the maximized contribution also weakly decreases. The welfare gain from extending operation over any given cost interval therefore weakly decreases. Every cutoff above the former $\beta^*$ gave a strictly lower payoff than $\beta^*$ before the change, since $\beta^*$ was the largest maximizing cutoff. Its payoff relative to that cutoff can only fall. Hence the optimal operating cutoff weakly decreases.

Set equilibrium quality to zero whenever the firm shuts down. The preceding comparisons then imply weakly lower quality provision throughout the cost support. By the firm's envelope condition \eqref{eq: minimum firm rent}, $\Pi(\beta)$ therefore weakly decreases.

Fix a cost type operating under both policies. For each consumer type, the expression $\phi(\theta)v(q)-\beta q$ is weakly decreasing in $q$ above $q^m(\theta,\beta)$. The quality reduction therefore weakly raises the integral in
\[
 U_0(\beta)
 =\int_\Theta
    [\phi(\theta)v(q^*(\theta,\beta))
       -\beta q^*(\theta,\beta)]\,dF(\theta)
    -\Pi(\beta).
\]
Together with lower firm profit, this implies that the lowest consumer's utility weakly increases.

Finally, for any $\theta'>\theta$, the consumer envelope gives
\[
 U(\theta',\beta)-U(\theta,\beta)
 =\int_\theta^{\theta'}v(q^*(t,\beta))\,dt.
\]
The quality comparison makes this difference weakly smaller. The change in consumer utility is therefore weakly decreasing in WTP.
\end{proof}

\subsection{Proof of Proposition~\ref{prop: fully flexible regulation}}
\label{app: fully flexible regulation}

\begin{proof}
I first obtain a welfare bound for arbitrary randomized mechanisms under Bayesian incentive compatibility. I then show that deterministic menus attain this bound and establish the comparison with monopoly.

\paragraph{The welfare bound and optimal qualities.} For a possibly randomized mechanism, let $q(\theta,\beta)$
denote the random quality allocated to a consumer reporting $\theta$ when the firm reports $\beta$. Expectations include the mechanism's randomization and the types of other consumers, who report truthfully. Consumer incentive compatibility requires $\int \mathbb E[v(q(\theta,\beta))]\,dG(\beta)$ to be nondecreasing in $\theta$. Applying the consumer envelope to interim utilities, and the firm envelope to expected profit, bounds the regulator's payoff by
\[
 \overline\omega
 \int_{\underline\beta}^{\overline\beta}\int_\Theta
 \mathbb E\!\left[
   \psi(\theta)v(q(\theta,\beta))
   -\widetilde\beta(\beta)q(\theta,\beta)
 \right]dF(\theta)\,dG(\beta).
\]

Let $\psi^I$ denote $\psi$ ironed with respect to $F$.\footnote{Formally, $\psi^I(F^{-1}(x))$ is the slope of the lower convex envelope of $x\mapsto\int_0^x\psi(F^{-1}(s))\,ds$.} The monotonicity of the consumer's interim allocation implies that replacing $\psi$ by $\psi^I$ weakly raises this bound. At each pair of types, expected virtual surplus cannot exceed the highest virtual surplus attainable by any quality. Hence the regulator's payoff is bounded above by
\[
 \overline\omega
 \int_{\underline\beta}^{\overline\beta}\int_\Theta
 \max_{z\geq0}
 \{\psi^I(\theta)v(z)-\widetilde\beta(\beta)z\}
 \,dF(\theta)\,dG(\beta).
\]
This also bounds randomized allocations, since no realization can exceed the displayed maximum.

Write $q(\theta,\beta)$ for these deterministic maximizers. They are nondecreasing in $\theta$ and constant on every interval where $\psi$ is ironed. The ironing inequality is therefore an equality for this allocation. Hence $q$ solves \eqref{eq: relaxed quality problem} without $q\geq q^m$. Moreover, $q$ is nonincreasing in $\beta$, since $\widetilde\beta$ is increasing.

Note that $\phi\leq\psi^I\leq\theta$. In particular, $\psi^I$ is positive near $\overline\theta$. The Inada condition then makes the maximum strictly positive for these consumers at every cost. So, the maximizing operating cutoff is $\overline\beta$: every cost type supplies positive quality.

\paragraph{Implementation.}
Set firm profit by \eqref{eq: minimum firm rent} and define
\[
 U_0(\beta)
 =\int_\Theta
   [\phi(\theta)v(q(\theta,\beta))
       -\beta q(\theta,\beta)]\,dF(\theta)
   -\Pi(\beta).
\]
The consumer envelope determines $U(\theta,\beta)$ and $p(\theta,\beta)=\theta v(q(\theta,\beta))-U(\theta,\beta)$. Since quality is nondecreasing in WTP, consumers prefer their assigned options at every cost. Consumer participation holds under the maintained assumption; condition~\eqref{eq: sufficient ir} is sufficient here as well.\footnote{Fix $\beta$. Since $q^m(\theta,\widetilde\beta(\beta)) \leq q(\theta,\beta) \leq q^e(\theta,\widetilde\beta(\beta))$, the same argument as in the main proof gives
\[
\begin{aligned}
 U_0(\beta)\geq{}
 \max_{z\geq0}
 \{\underline\theta v(z)-\widetilde\beta(\beta)z\}
 +[\widetilde\beta(\beta)-\beta]
 \int_\Theta q(\theta,\beta)\,dF(\theta)-\Pi(\beta).
\end{aligned}
\]
The maximum is strictly positive by the Inada condition. Moreover, quality decreases in cost, so $\Pi(\beta)\leq(\overline\beta-\beta) \int_\Theta q(\theta,\beta)\,dF(\theta)$. So $U_0(\beta)\leq0$ would require $\widetilde\beta(\beta)<\overline\beta$. But then the maximum is at least the right-hand side of \eqref{eq: sufficient ir}, while $\Pi(\beta)$ is bounded by its left-hand side, as in the main proof. Since $\widetilde\beta(\beta)\geq\beta$, the displayed inequality implies $U_0(\beta)>0$, a contradiction.}

Offer the firm the resulting menus, without allowing additional options. If type $\beta$ chooses the menu intended for $\hat\beta$, its profit is
\[
 \pi(\hat\beta;\beta)
 =\Pi(\hat\beta)
   +(\hat\beta-\beta)
      \int_\Theta q(\theta,\hat\beta)\,dF(\theta)
 \leq\Pi(\beta).
\]
The inequality follows from the firm envelope and the fact that total quality is nonincreasing in cost. Firm participation follows from $\Pi\geq0$. These menus therefore attain the welfare bound, including the bound over randomized Bayesian incentive-compatible mechanisms.

\paragraph{Comparison with monopoly.}
Fix $\beta$ and a consumer type $t>\theta^*(\beta)$. By the definition of $\theta^*(\beta)$, there is $s<t$ satisfying the inequalities in \eqref{eq: baseline cutoff}. In particular, every average of $\psi$ over an interval ending at $s$ is at most $(\widetilde\beta(\beta)/\beta)\phi(s)$. Taking the lower endpoint to $s$ gives the same bound on $\psi(s)$.

The proof of Theorem~\ref{thm: baseline regulation} shows that $\psi-\phi$ is nonincreasing. Hence, for $\theta\in[s,t]$,
\[
 \psi(\theta)
 \leq \phi(\theta)
     +\left(\frac{\widetilde\beta(\beta)}{\beta}-1\right)\phi(s)
 \leq \phi(t)
     +\left(\frac{\widetilde\beta(\beta)}{\beta}-1\right)\phi(s).
\]
Splitting intervals at $s$ therefore bounds every average of $\psi$ over an interval ending at $t$ by this last expression. The ironed value at $t$ cannot exceed the largest of these averages: inside a pooling interval, the average from its left endpoint to $t$ is at least the ironed value; otherwise, let the lower endpoint approach $t$. Consequently,
\[
 \begin{aligned}
 \psi^I(t)
 \leq
 \sup_{\underline\theta\leq a<t}
 \frac{\int_a^t\psi(\theta)\,dF(\theta)}{F(t)-F(a)} \leq \phi(t)
     +\left(\frac{\widetilde\beta(\beta)}{\beta}-1\right)\phi(s)
 <\frac{\widetilde\beta(\beta)}{\beta}\phi(t).
 \end{aligned}
\]
The last inequality is strict because $\widetilde\beta(\beta)>\beta$ and $\phi(t)>\phi(s)$. Also, $\phi(s)\geq0$, so $q^m(t,\beta)>0$. At this monopoly quality, the marginal payoff $\psi^I(t)v'(q^m(t,\beta))-\widetilde\beta(\beta)$ is strictly negative. Since $\psi^I(t)\geq\phi(t)>0$, strict concavity implies $q(t,\beta)<q^m(t,\beta)$, as claimed.
\end{proof}

\end{appendix}

\bibliography{literature_references.bib}
\bibliographystyle{econ}
\addcontentsline{toc}{section}{Reference}

\end{document}